\def\beq{ \begin{equation}}
\def\eeq{ \end{equation}}
\def\ba{ \begin{eqnarray}}
\def\ea{ \end{eqnarray}}

\def\half {{1\over 2}}
\documentstyle[aps,preprint,12pt,epsf]{revtex}
\tightenlines

\title{ Acceleration Radiation for Orbiting Electrons.\footnote{To be published in The Proceedings of the Quantum Aspects
 of Beam Physics, Monterey, Jan 1998,  ed Pisin Chen,} }

\author{ W. G. Unruh}
\address{ Program in Cosmology and Gravity of CIAR\\Dept 
Physics and Astronomy\\University 
of B.C.\\ Vancouver, Canada V6T 1Z1}
\begin{document}
\maketitle

\begin{abstract}
This paper presents an analysis of the radiation seen by an observer
in   circular acceleration, for a magnetic spin. This is applied to an
electron in a storage ring, and the subtilty of the interaction of the
spin with the spatial motion of the electron is explicated. This
interaction is shown to be time dependent (in the radiating frame),
which explains the strange results found for the electron's residual
polarisation in the literature.  Finally, some brief comments about the
radiation emitted by an accelerating detector are made where  it is
shown that the spectrum is correlated in that particles are emitted in
pairs. 
 \end{abstract}

A uniformly accelerated objects sees the vacuum fluctuations as a
thermal bath\cite{unruh}. For an acceleration which is constant in
amplitude and direction, this temperature is given by 
\beq
 T= {a\over 2\pi} \left({
\hbar \over c k} \right)
\eeq
 Since it requires an acceleration of $2.6\cdot10^{22}$ cm/sec$^2$ to
produce a temperature of $1K$ the experimental verification of this
prediction is difficult, although recent advances in the laser
acceleration of electrons promise to produce accelerations of over a
100 times this value\cite{laser}. These would give effective
temperatures of the order of room temperature, but for very short
times. Those brief intense accelerations make it difficult to measure
the effects of the thermal bath on the object.

 As was pointed out by Bell
and Leinaas\cite{BL}, the spin of an electron in a magnetic field is a
possible candidate for a detector of acceleration radiation. Such an
electron has two energy levels, and an examination of the population of
those two energy levels after a long time can be used to measure the
bath that the electron sees itself in at the Larmour precession
frequency.

   Since even room temperature corresponds to very low frequencies(
compared with the compton frequency of the electron), and since dipole
radiation is suppressed by the third power of the frequency,
 the use of the electron spin as a detector of the acceleration bath
requires a very long period of acceleration to produce an effect.
 Bell and Leinaas\cite{BL} therefore suggested that circular
accelerations (placing the electron into a circular orbit) be used
instead. In an electron storage ring, the
acceleration experienced by the electrons in the bending magnets
corresponds to high temperature(about $10^5$K for HERA), and relatively
short decay times (about half an hour).The electron can be kept in the
storage ring for that time and be expected  to  equilibrate with the
thermal  bath. They thus suggested that the  residual polarisation of
the electrons   is therefore a measure of the temperature of the bath
seen by the electrons.      Jackson\cite{Jackson-comment} however has
raised serious doubts about the use of a thermal model to explain the
residual polarisation of the electrons in such storage rings. In
particular, he has argued that the details of the polarisation of the
electron as a function of the $g$ factor found by Derbenev and
Kondratenko\cite{DK} (hereafter DK) and himself\cite {Jackson} as
plotted in Figure 1, made such a thermal explanation highly suspect.
Zero polarisation does not occur at zero coupling with the field by the
spin (ie, zero g factor) nor at zero anomalous coupling ($g=2$) but at
about $g=1.2$.  Furthermore, there are details in the polarisation
curve which do not resemble what one would expect from the
equilibration with a thermal bath.

\epsfysize=3in
\centerline{\epsfbox {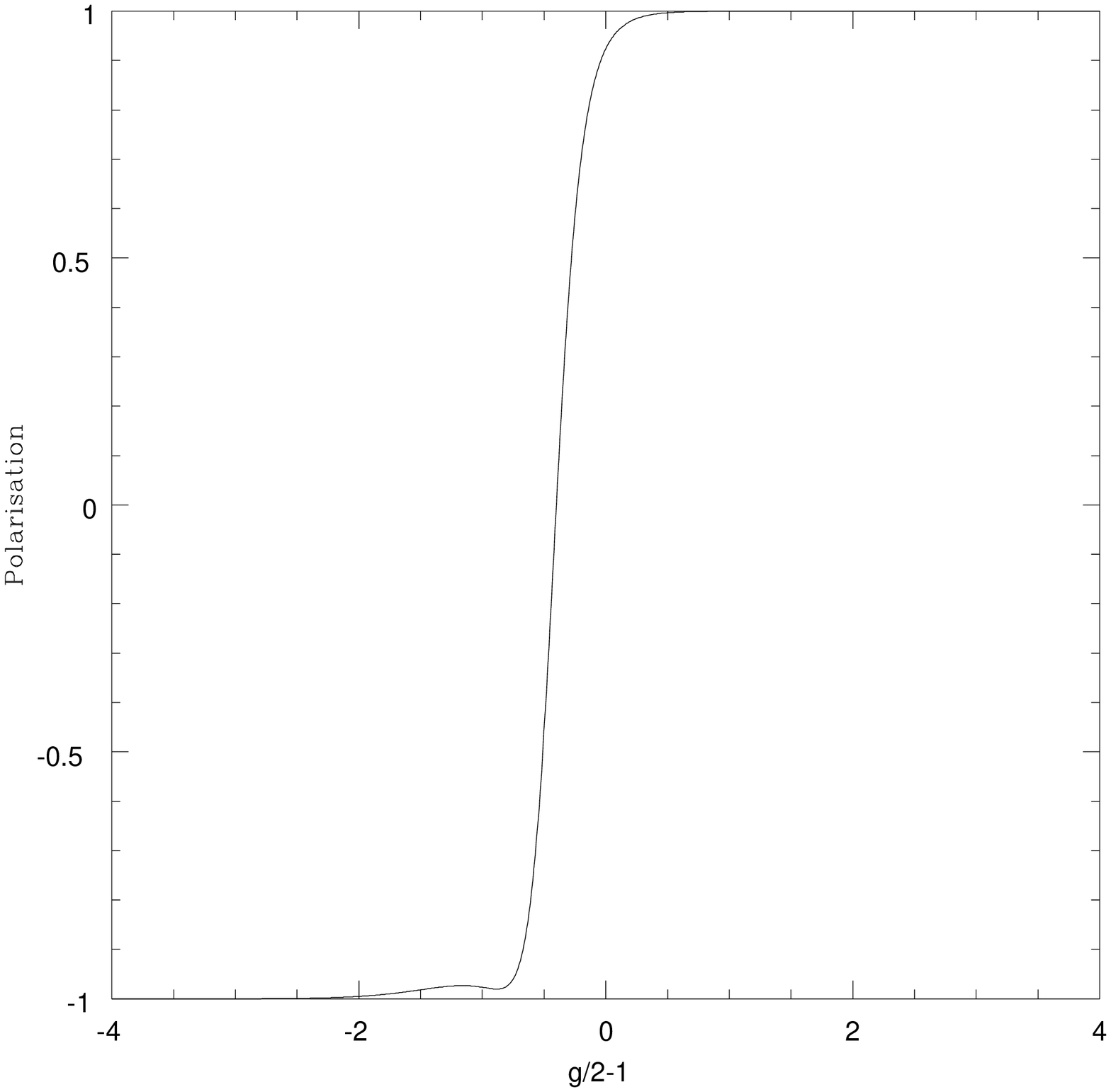}}
\centerline{Figure 1}
\centerline{\it Polarisation of an electron vs the $g$ factor }
\bigskip

The primary purpose of this paper is to analyse the behaviour of an
electron undergoing circular acceleration. I will restrict myself to
the electron's travelling at constant velocity in a constant magnetic
field (ie, I assume that some tangential electric  field is present to
make up for the energy lost in synchrotron radiation).  My conclusion
is that the electron does respond to a thermal bath, but one with a
frequency dependent temperature.  
 However, the electron is actually a system with two field detectors,
which are furthermore coupled to each other. In addition to the spin
there are the vertical fluctuations in the orbit. A simple coupling
should present no difficulties for understanding the system,
 but in this case that
coupling is also time dependent (in the radiating frame of the spin).
This time dependent coupling distorts the response of both detectors.
( Almost all of the results here was already
in
the paper of  Bell and Leinaas \cite{BL}(hereafter BL), at least implicitly.
I hope that this different exposition may make some of the points clearer)

Finally, the emission of radiation by an accelerated detector
 (coupled to a scalar
field in this case) is very briefly analysed and it is shown that the
emitted radiation comes out in correlated pairs of particles. This
section is brief, and more detailed investigation of these correlations
will be reported elsewhere.

Throughout I will use the convention that $\hbar=c=1$.

\section{Pole structure of the fluctuations}

The fluctuations in the electromagnetic (and other massless fields) can
be characterised by the two point correlation function. Because the
vacuum state is a Gaussian state, the two point correlations completely
characterise the field.  These fluctuations typically have a term in
the denominator of the form
\beq
D= (\Delta t)^2 - (\Delta x)\cdot(\Delta x)
\eeq
where $\Delta t=t-t'$, the two times in the
 correlation function, and similarly for $\Delta x$.
 For circular acceleration, this term becomes
\beq
D= \gamma^2(\tau-\tau')- 4R^2\sin(\half\gamma\Lambda(\tau-\tau'))^2
\eeq
where $\tau$ is the proper time along the circular path, $R$ the radius
of the path, $\Lambda$ is the angular frequency (in the lab frame) of the orbit,  and  the relativistic gamma factor is given by
\beq
\gamma^2={1\over (1-R^2\Lambda^2)}. \label{gamma}
\eeq
 In calculating the response of the a
detector at its resonant frequency, the fourier transform of this
correlation function will be important, and thus the zeros of $D$,
which will become poles of the response function, determine the
response\cite{KSY}. I will be interested in the case of extreme relativistic
motion$\gamma>>1$. The poles of the correlation function will
correspond to zeros of $D$.

The most obvious zero is at $\tau=0$, but others 
  are scattered about the complex
$\tau$ plane.  There is another pole near
 $\tau=\tau'$. Expanding $\sin$ in a taylor series, I get
\beq
\gamma^2\tau^2- 4R^2\sin( \half\Lambda\gamma\tau )^2\approx \tau^2 
+ {8\over 6\cdot 16}R^2 \gamma^4\Lambda^4\tau^4 +...
\eeq
which has a zero at 
\beq
\tau=\pm\tau_1\approx\pm i {2\sqrt{3} \over \sqrt{\gamma^2-1}\Lambda}
\eeq
where I have use eqn \ref{gamma} to eliminate $R$.
Figures 2 and 3 graphically solve the equation
\beq
(x+iy)^2= {{\gamma^2-1}\over \gamma^2}\sin^2(x+iy)
\eeq
or
\ba
x&&= \sqrt{\gamma^2-1\over \gamma^2} \sin(x) \cosh(y)\\
y&&=\sqrt{\gamma^2-1\over \gamma^2} \cos(x) \sinh(y)
\ea
{\rm and}
\ba
x&&= -\sqrt{\gamma^2-1\over \gamma^2} \sin(x) \cosh(y)\\
y&&=-\sqrt{\gamma^2-1\over \gamma^2} \cos(x) \sinh(y)
\ea

\epsfysize=3in
\centerline{\epsfbox{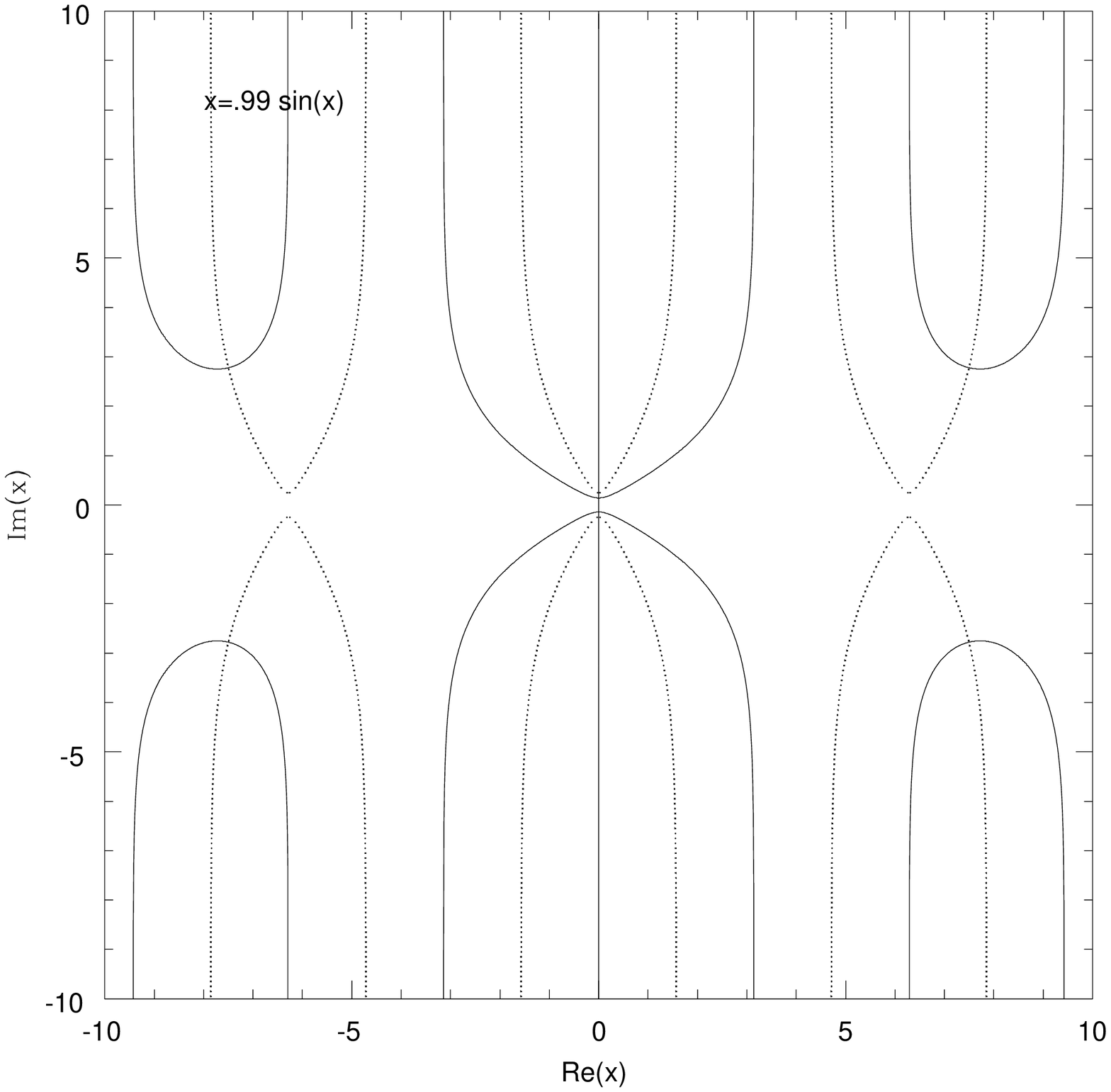}}
\centerline{Figure 2 }
\centerline{\it Real and imaginary parts of x=.99 sin(x) }
\bigskip

\epsfysize=3in
\centerline{\epsfbox{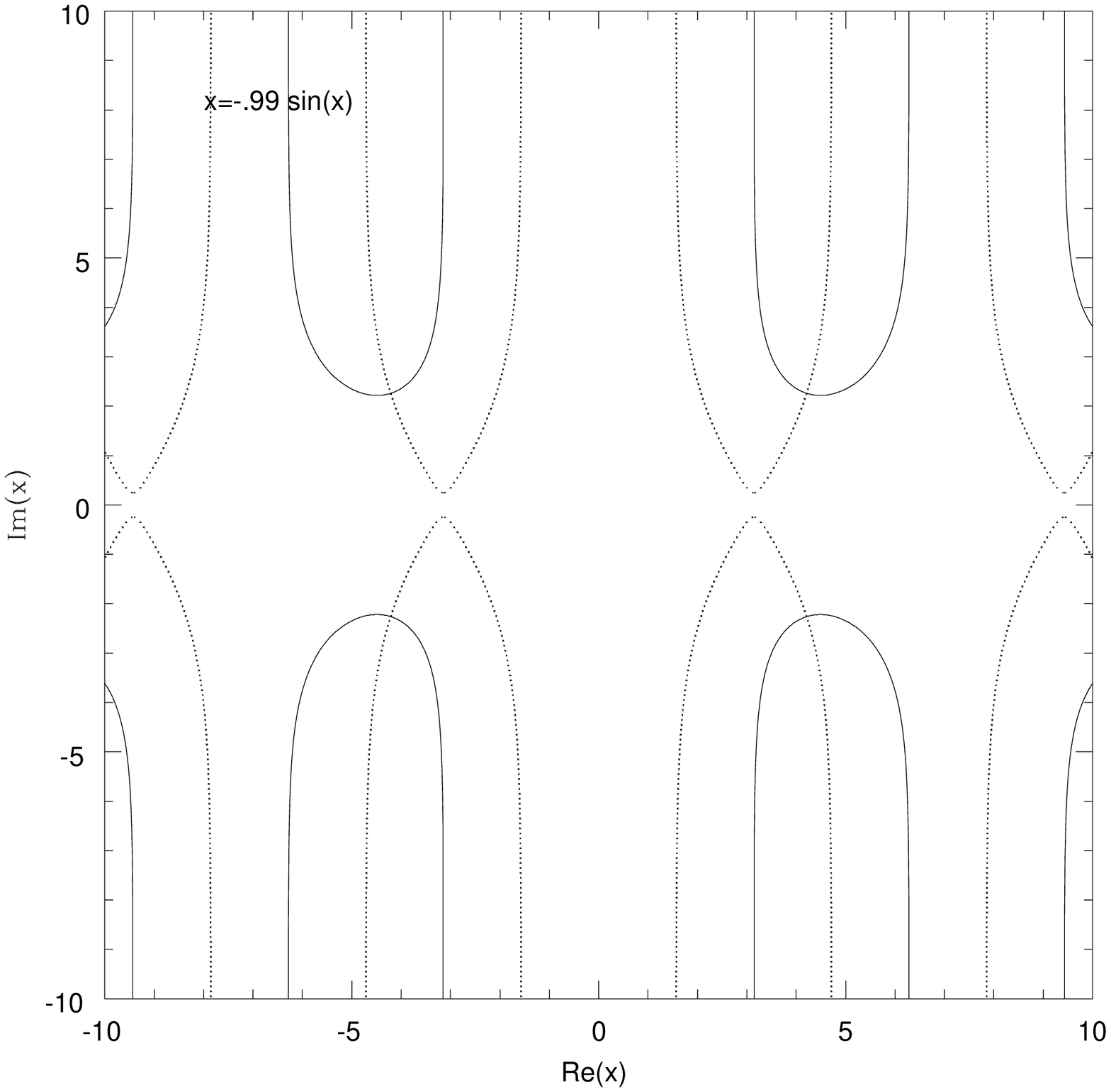}}
\centerline{Figure 3}
\centerline{\it Real and imaginary solutions of x=-.99sin(x) }
\bigskip

The other zeros of the equation (which give poles in the correlation
function)
 all have a value of $x$ and $y$  of order unity or larger, which are
 of order  $\gamma$ larger in magnitude than the pole at $x=0,
~y\approx{\sqrt{3}\over \gamma} $. Their contribution to the Fourier
transform of the correlation function
  will be very small for large $\gamma$ unless $\omega$ is very near
zero.   

The fluctuations are Gaussian. If they are also stationary (depending
only on $\tau-\tau'$), then the fluctuations can be regarded as
thermal, with a possibly frequency dependent temperature. As BL showed,
for a detector of massless scalar radiation in circular orbit , the
temperature of the fluctuations is given by figure 3. The temperature
is frequency dependent, and is scattered around the value expected from
equation 1 for linear acceleration.  However, just because the
temperature is not frequency independent, does not make it act any the
less like a temperature especially if the detector responds only to a
narrow range of frequencies. The more important aspects are the lack of
correlation between the fluctuations at the various frequencies (which
comes from the stationarity of the correlation function) and the
gaussian nature of those fluctuations.

For a scalar detector in circular orbit, the effective temperature of
the scalar field fluctuations turns out to be
 
\beq
T(\omega)= {\omega \over \ln\left(  1+ {4\sqrt{3}\omega  \over a_c } e^{ 2\sqrt{3}\omega \over a_c}\right) }
\eeq
Thus, for low $\omega$, the temperature is 
\beq
T\approx {a\over 4\sqrt{3}}
\eeq
while for high $\omega$, it is approximately
\beq
T\approx {a_c\over 2\sqrt{3}}
\eeq
Figure 4 gives a plot of $T(\omega)/a$ vs $\omega/a_c$.   The straight
  line is the temperature for the same acceleration but with
linear rather than circular acceleration.

\epsfysize=3in
\centerline{\epsfbox{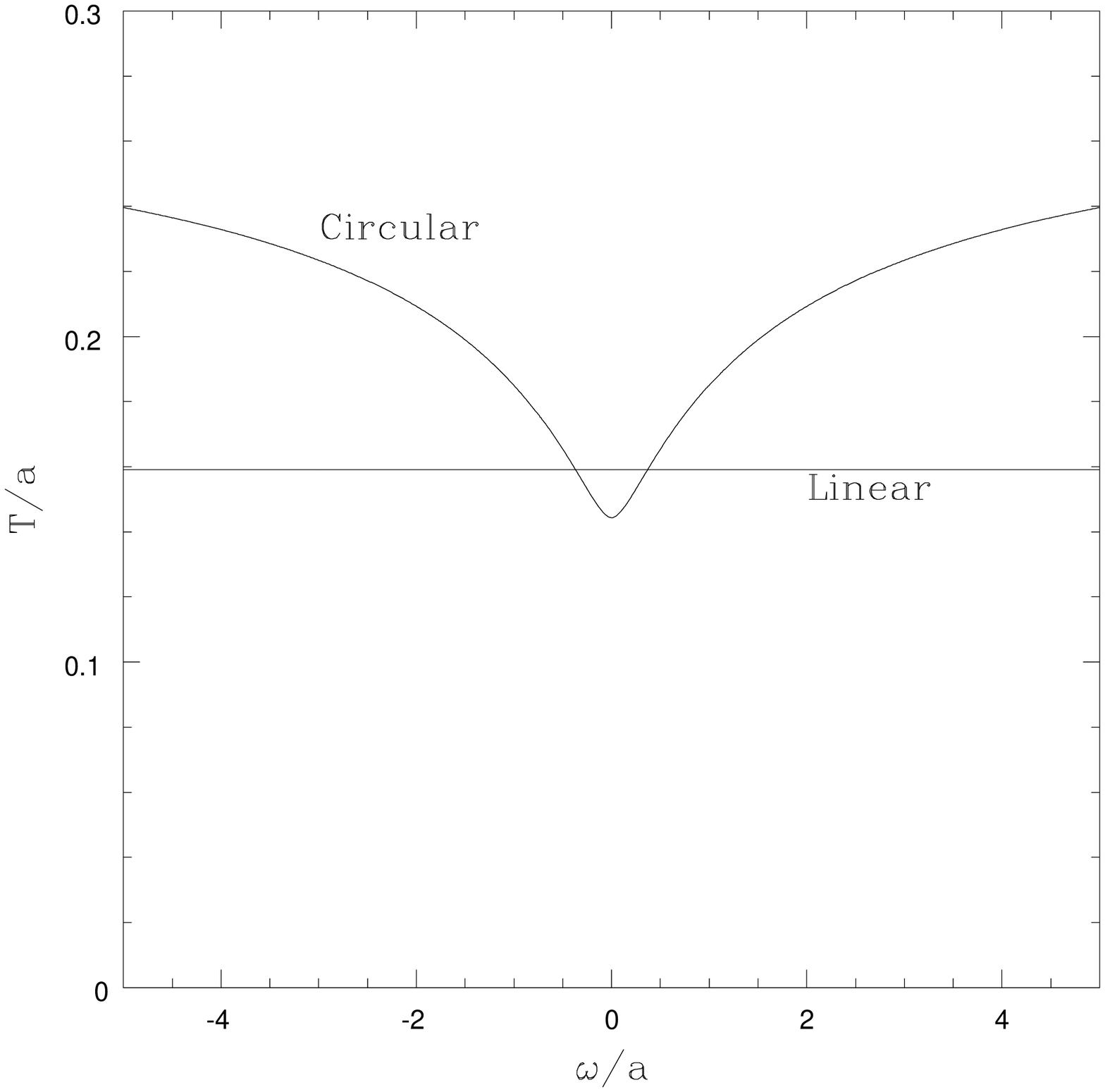}}
\centerline{Figure 4}
\centerline{\it The Temperature seen by a circular and 
linear accelerated scalar field detector }
\bigskip

This was calculated on the basis of using the residue of only the three
poles closest to zero. The contribution of the neglected poles is small
and I estimate it to be less than about  .2\% for $\gamma$ of 10.

For a scalar detector the estimation of the response is now simple. The
detector is in contact with a thermal bath, and at long times, one
would expect it to come into equilibrium with that bath with the same
temperature $T$ describing its internal state. This is born out for a
scalar detector (where for example the response of a harmonic
oscillator like internal detector can be solved exactly). If the
coupling becomes too strong, then of course the internal state is no
longer at the naive temperature due to correlation effects between the
bath and the detector\cite{UZ}. However all situations of interest to
us will be well within the weak coupling regime.

\section{Electromagnetism}

Jackson\cite{Jackson-comment} has criticized the use of the equilibrium
polarisation of the spin of an electron in a storage ring as a detector
of the thermal nature of the field as seen by an accelerated observer.
Even Bell and Leinaas in their second paper raise doubts about  
 regarding the residual polarisation as a measure of
the temperature  of the thermal bath seen by a circularly accelerating
electron.  The distribution amongst the states of the spin simply does
not look thermal, given the interaction of the spin with the field. 
However they
do not explain why the detector response differs from the naive explanation.
The expectation that the spin polarisation should simply have
a distribution which corresponds to the thermal distribution of the
field is too naive. One must understand both the interaction of the
spin with the bath and with other degrees of freedom of the system
itself. After all, the snow seen on the screen of a TV set has a
temperature vastly higher than the room temperature fluctuations which
cause it, due to the interaction with other components (eg, amplifiers,
etc). The fact that the observed temperature is so much higher than the temperature
which causes the snow is
not a failure of the TV set to act as a detector of thermal
fluctuations, but rather a failure to understand how the TV set works.
As we shall see, the electron also has similar complex behaviour
including time dependent couplings which act as ``amplifiers".

  The first question to ask is whether or not the electromagnetic field
appears to be a thermal field to the circularly accelerating observer.
Clearly the fluctuations are Gaussian, since they are simply the vacuum
fluctuations of the electromagnetic field, which are Gaussian. Thus the
key question is whether or not they are stationary. Ie, is the two-point
correlation function of the fields a function only of the difference in
times, or does it also depend on the absolute time?

For the scalar field this question is easily answered in the
affirmative, but   the electromagnetic field, being a vector field, has
the additional complication of needing to have the components one is
correlating be specified first.  Ie, even for a static observer in the
vacuum, the correlation function $<e(t)_k B(t)^k e(t')_j B(t')^j>$ is
not stationary for arbitrary time dependent unit vector $e_k$. In the
case of the circular acceleration case as there are
numerous frames  in which one could calculate the correlation
 function between the components of the electromagnetic field.  Three in
particular spring to mind, namely the inertial system of basis vectors (ie,
parallel transported along the curve), the rotating system (which
rotates at the same rate as the particle revolves around the circle) or
the Fermi-Walker transported system. Since each of these sets of basis
vectors rotates with respect to the other, one would expect that only one
 of them will be the
one (if any) in which the fluctuations are stationary. The answer is that
it is the basis set which co-rotates with the particle. One can define a
Killing vector field, a sum of time translation and rotation
\beq
\zeta^\mu = (1,-\Lambda y,\Lambda x,0)
\eeq
such that the particle follows one of the integral curves of this
Killing vector field. It will then be the components of the
electromagnetic field as Lie transported along this field which will be
stationary. (Since the vacuum state is
invariant both under time translations and rotations,   it is also
invariant under this Killing transformation). However, the electron
effectively couples to only a few of these components. It will thus
turn out for the electron that it is the correlation functions in the
FW frame which are important, and the particular components of
importance will again be stationary in that frame.

The correlation function for the electromagnetic
 field in Minkowski coordinates is
\ba
<F(t,x)_{\mu\nu}F(t',x')_{\rho\sigma}> 
&&= 4\partial_{[mu}\eta_{\nu][\sigma}\partial_{\rho'} 
{1\over \pi \left( (t-t')^2-(x-x')^2\right)}\\
&&= 32{\Delta x_{[\mu}\eta_{\nu][\sigma} \Delta x_{\rho]}
 \over \pi (\Delta x_\alpha \Delta x^\alpha)^3}- 8 {\eta_{\mu[\rho} 
\eta_{\sigma] \nu}\over \pi(\Delta x_\alpha\Delta x^\alpha)^2}
\nonumber
\ea
where  the square brackets around indices indicates
anti-symmetrisation. $\Delta x^\mu$ is the four-vector $ [t-t',x-x']$.
Define a set of co-rotating basis vectors appropriate to the uniformly
circling particle
\ba
e_\tau^\mu&&=\gamma[1,-R\Lambda \sin(\Lambda\gamma\tau)
           ,R\Lambda \cos(\Lambda\gamma\tau),0]
\nonumber\\
e_\rho^\mu&&= [0,\cos(\Lambda\gamma\tau)
         ,\sin(\Lambda\gamma\tau),0]\nonumber
\nonumber\\
e_\phi^\mu&&=\gamma[R\Lambda, -\sin(\Lambda\gamma\tau)
        , \cos(\Lambda\gamma\tau),0]\\
e_z^\mu&&=[0,0,0,1]\nonumber
\ea
where $e^\mu_\tau$ is the tangent vector to the curve the particle is
following, I get for the correlation functions of interest
\ba
<B_\rho(\tau) B_\rho(0)>&&=<E_\rho(\tau) E_\rho(0)>   \\
&&= {4\gamma^2\over \pi D^3}
\left( \cos(\gamma\lambda\tau)(\gamma^2\tau^2 -2R^2(1+\lambda^2R^2)) \right.
\nonumber\\
&&\quad \left.  -4\Lambda R^2\gamma\tau \sin(\Lambda\gamma\tau)
+R^2( 2+2\Lambda^2R^2 +\Lambda^2\gamma^2\tau^2)\right)
\nonumber
\\
<B_\rho(\tau)B_\phi(0)>&&=<E_\rho(\tau)E_\phi(0)>  \\
&&={4\gamma^2\tau\over \pi D^3}\left(  -2\Lambda R^2+2\Lambda R^2 \cos(\Lambda\gamma\tau)  
  +\gamma\tau\sin(\Lambda\gamma\tau)\right) 
 \nonumber
\\
<B_\phi(\tau)B_\phi(0)>&&=<E_\phi(\tau)E_\phi(0)={4\over \pi D^3}(\cos(\Lambda\gamma\tau)(\gamma^2\tau^2 +2R^2)-2R^2)
\\
<B_z(\tau)B_z(0)>&&=<E_z(\tau)E_z(0)\\
&&={4\gamma^2\over \pi D^3} 
\left(\cos( \Lambda\gamma\tau)
(\gamma^2\tau^2 \Lambda^2R^2 -2R^2-2\Lambda^2R^4)\right. 
\nonumber\\
&&\qquad\left. -4\sin(\Lambda\gamma\tau) \gamma\tau\Lambda R^2
 +\gamma^2\tau^2 +2\Lambda^2R^4 +2R^2\right)
\nonumber
\\
<B_\rho(\tau)E_z(0)> &&= -<B_z(\tau)E_\rho(0)>=-{4R\gamma^2\over \pi D^3}
 \left( \cos(\Lambda\gamma\tau)(\Lambda\gamma^2\tau^2 -4\Lambda R^2) \right. \\
&&\qquad \left. -2\gamma\tau\sin(\Lambda\gamma\tau)
(1+\Lambda^2R^2) + \Lambda(\gamma^2\tau^2 + 4R^2)\right)
\nonumber\\
<B_\phi(\tau)E_z(0)>&&= <B_z(\tau)E_{\phi}(0)>
={4R\gamma^2\tau\over \pi D^3} \left( 2\cos(\Lambda\gamma\tau) 
+\sin(\Lambda\gamma\tau) \Lambda\gamma\tau -2\right)
\ea
All correlations   are zero, except those which 
can be derived from these by the symmetry, $<A(\tau)B(0)>=<B(-\tau)A(0)>$.
 These correlations functions are thus much more
complex than those for the uniformly accelerating observer, or a
stationary observer with the field in a thermal state, where all cross
correlations are zero, and all diagonal correlations are the same as
each other.

Let us now examine the equation of motion of the electron, and its
interaction with the field fluctuations along the path of the particle.
I will attack the problem in three steps. Firstly I will examine the
equations of motion of the spin on its own, assuming that the electron
follows the circular orbit exactly (ie, all fluctuations from this path
will be ignored.) Then I will include the effects of fluctuations in
the path in the vertical (z) direction. This will introduce another
internal  system, a harmonic oscillator in the $z$ component of the
motion. Oscillations in the other two directions turn out not to couple
to the spin system to lowest order, and will be ignored. I will calculate 
the    the harmonic oscillations in the z direction driven by the electromagnetic field  and the effects of such oscillations on the
 spin. Those effects arise because of the changes in the Thomas 
precession of the spin induced by this extra motion.

The motion of the spin of the electron has been extensively studied.
  The electron spin, in the absence of an magnetic field in the rest
frame of the spin, is Fermi-Walker transported along the path of the
particle.  Fermi-Walker transport for a vector which is spatial in the
rest frame of the particle is parallel transport continually projected
back to the spatial subspace (ie orthogonal to the velocity vector).
 (In general parallel transport along a non-geodesic will take a vector
orthogonal to the velocity vector and create components parallel to the velocity. Fermi-Walker transport continuously subtracts that velocity-parallel
component from the motion of the particle.) Formally it is defined for
a vector $w^\mu$ which is orthogonal to the velocity of the particle
$V^\mu$ as
\beq
{D_{FW}w^\mu\over D\tau}=
 {D w^\nu\over D\tau}(\delta_\nu^\mu 
- V_\nu V^\mu)={D w^\mu\over D\tau}+ V^\mu a_\nu w^\nu
\eeq
where $V^\mu$ is the tangent vector to the curve the particle follows,
and $a^\mu$ is the acceleration vector $DV^\mu\over D\tau$. Note that
this equation maintains the relation $V_\mu w^\mu=0$, ie the vector
$w^\mu$ remains  spatial in the rest frame of the particle. The presence
of a magnetic field in the rest frame of the particle exerts a torque
on the spin. In the circular accelerating system, the spin equation is
simply\cite{JacksonEM}

\beq
{D_{FW}s^\mu\over D\tau}=  g\mu_0 F^{\rho\nu}s_\nu \left(\delta^\mu_\rho -V^\mu V_\rho\right)
\eeq
or, explicitly expanding the Fermi-Walker derivative for the circularly accelerating electron, and the definition of the components $s_i$ in the co-rotating frame
\beq
{ds_i\over d\tau} = \epsilon_{ijk} (g\mu_0 {\bf B}^j- \gamma^2\Lambda e_z^j) s^k 
\eeq
where g is the ``g" factor for the spin, and $\mu_0$ classical dipole
moment $e\over 2m$, and $\epsilon_{ijk}$ the antisymmetric tensor for 
the spatial indices, such that $\epsilon_{123}=1$. Spatial indices are 
raised and lowered by the 4-metric and thus an extra minus sign tends
 to appear in the formulas due to my sign convention for the 4-metric.
 The external field portion of  ${\bf B}_j$ is the constant
field (in the rest frame of the electron) $\gamma B_{0}e_{zj}$ plus the vacuum fluctuations $B_j(\tau)$
whose correlations  are calculated above.  Note that I have chosen the
constant $B_0$ field (in the inertial frame) to lie along the  $z$
axis. If we assume, as I will, that the same magnetic field is
responsible for the circular trajectory of the particle, then
\beq
\Lambda = {e B_0\over m\gamma}=2\mu_0 B_0/\gamma
\eeq

 These equations of motion can be derived from a Hamiltonian. Taking
 $s^i=\half\sigma^i$, the Pauli spin matrices, the Hamiltonian is
\beq
H= g\mu_0\half\sigma\cdot {\bf B} - \gamma^2\Lambda \sigma_z
\eeq
where {\bf B} is the total magnetic field as seen in the rest frame of
the electron.  This is composed of the fixed field $\gamma B_0$ in the
z direction, and the fluctuating vacuum field, $B^i$.  To calculate the
equilibrium population, I use the standard technique of   time
dependent perturbation theory to calculate the transition probabilities
of the spin going from one state to the other.  Defining,

 \ba
\sigma_+=\half(\sigma_\rho+i\sigma_\phi)\\
\sigma_-= \half(\sigma_\rho+i\sigma_\phi)\\
B_+(\tau)= \half(B_\rho(\tau) +iB_\phi(\tau))\\
B_-(\tau)=\half(B_\rho(\tau) -iB_\phi(\tau))
\ea
we have
\beq
H= \half\left((g\mu_0(\gamma B_0+B_z) 
-\gamma^2\Lambda) \sigma_z +g\mu_0 (B_+\sigma_- +B_- \sigma_+)\right)
\eeq
Defining 
\beq
\Omega= g\mu_0\gamma B_0-\gamma^2\Lambda= (g-2)\mu_0 \gamma B_0
\eeq
  and noticing that the
component $B_z$ of the vacuum field in the $z$ direction will not (to
lowest order) affect the transition probability between the two levels
of the spin in the effective $\Omega$ field, the Hamiltonian becomes
\beq
H= \half \Omega \sigma_z  + B_+\sigma_- + B_- \sigma_+
\eeq
Using first order time dependent perturbation theory, the transition
probability per unit time to go from the lower to the upper state or
the upper to the lower state is
\ba
P_\uparrow&&= Lim_{T\rightarrow\infty}{1\over 2T}\Sigma_{<1|}  | \int_{-T}^T < 1,+|B_-(\tau)|0,->d\tau  |^2\\
&&=Lim_{T\rightarrow\infty}
 {1\over 2T} \int_{-T}^T\int_{-T}^T <0| B_+(\tau)B_-(\tau')|0>  
 e^{-i\Omega(\tau-\tau')}d\tau d\tau'\\
&&=\int<B_+(\tau-i0)B_-(0)>e^{-i\Omega\tau} d\tau
\ea
where the last line has used the stationarity of the correlation function.
 Also   the symmetry of the electromagnetic correlation function 
$<B_a(\tau+i0)B_b(0)>= <B_b(0)B_a(t-i0)>=<B_b(-\tau+i0)B_a(0)>$ I also get
\beq
 P_{\downarrow}= \int <B_+(\tau+i0)B_-(0)>e^{-i\Omega\tau} d\tau
\eeq
Assuming that the up conversions, and the down conversions are 
uncorrelated (which the long time scale of the decay process may
 make a good approximation), 
the ratio of these is the ratio of the equilibrium populations of
 the two levels.
\beq
{N_{up}\over N_{dn}}= {P_{\uparrow}\over P_{\downarrow}}
\eeq
 
 I get that 
\ba
<B_+(\tau)&&B_-(0)>= {1\over 4}(<B_\rho(\tau)B_\rho(0) 
+B_\phi(\tau)B_\phi(0) -i (B_\rho(\tau)B_\phi(0) 
-B_\rho(-\tau)B_\phi(0))>)
\nonumber\\
&&= {\gamma^4\over \pi D^3}\left( \cos(\Lambda\gamma\tau)\left[2\tau^2 -\tau^2\Lambda^2R^2 +4\Lambda^4 R^6-4\Lambda^2R^4 +i(4\Lambda^3R^4\tau  -4\Lambda R^2\tau)\right] \right.
\nonumber\\  
&& \qquad \left. -{2\tau\over \gamma} \sin(\Lambda\gamma\tau)\left[  2\Lambda R^2 +i\tau\right]  +{4\Lambda^2 R^4\over \gamma^2} +\Lambda^2 R^2\tau^2 +i{4\Lambda R^2\tau\over \gamma^2}  \right) 
\ea
in the co-rotating frame. Note that this particular correlation
function, unlike the general electromagnetic correlation function, is
stationary  in any  frame which rotates uniformly
 about the $z$ axis. Under a transformation to another
rotating frame (around the $z$ axis) the $B_+(\tau)$ field goes to
$e^{i\nu\tau} B_+(\tau)$ while $B_-(\tau)$ goes to $e^{-i\nu\tau}$ Thus
the correlation function goes to $<B_-(\tau)B_+(\tau')> \rightarrow
e^{-i\nu(\tau-\tau')} <B_-(\tau-\tau')B_+(0)>$ which is again
stationary. One therefore expects to find a ``thermal" distribution
 in all frames. However, this distribution will have a 
temperature which will be wildly varying except in one 
particular frame. This will play a role in the interpretation of the results
presented below.

 For large $\gamma$, just as for the scalar case, the dominant poles
 in the denominator are again at
$\tau=0$ and at $\tau=\pm i {2\sqrt{3}\over \gamma^2\Lambda}$. 
However the order of these poles changes. Since
the numerator is quadratic in $\tau$ at $\tau=0$, the former pole is a
quadratic pole while the latter poles are triple poles. Evaluating the
residue at these poles, and keeping terms only to lowest order in
$1/\gamma$, I find
\ba
P_\uparrow=&& 4(W+1)^2 \mu_0^2(\gamma^2\Lambda)^3  
\left(\left[F_1(W) -{|\Omega|\over \Omega}F_2(W)\right]e^{-2\sqrt{3} |W|}
-\Theta(-W) F_0(W)\right)
 \label{trans}\\
P_{\downarrow}=&&4(W+1)^2 \mu_0^2(\gamma^2\Lambda)^3 
\left(\Theta(W) F_0(W) +\left[F_1(W) 
-{|W|\over W}F_2(W)\right] e^{-2\sqrt{3} |W|}\right)\nonumber
\ea
where $W=\Omega/\Lambda\gamma^2=\Omega/a= g/2-1$.  The common factor of $4(W+1)^2\mu_0^2$ term is just $(g\mu_0)^2$.
 The functions $F_i$ are given by
\ba
F_0= {2\over 3}(W+1)((W+1)^2+1)  \\
F_1= {\sqrt{3}\over 288}  \left( 84 W^2-168 W +133\right)\\
F_2= {1\over 288} \left( 144 W^2-294 W +192\right)\\
\ea
In figure 5 I have plotted the polarisation 
\beq
{\cal P}= {P_\uparrow-P_\downarrow \over P_\uparrow+P_\downarrow}
\eeq
vs the normalised frequency $W$, or the anomalous  $g$ factor $g/2-1$.  

\epsfysize=3in
\centerline{\epsfbox{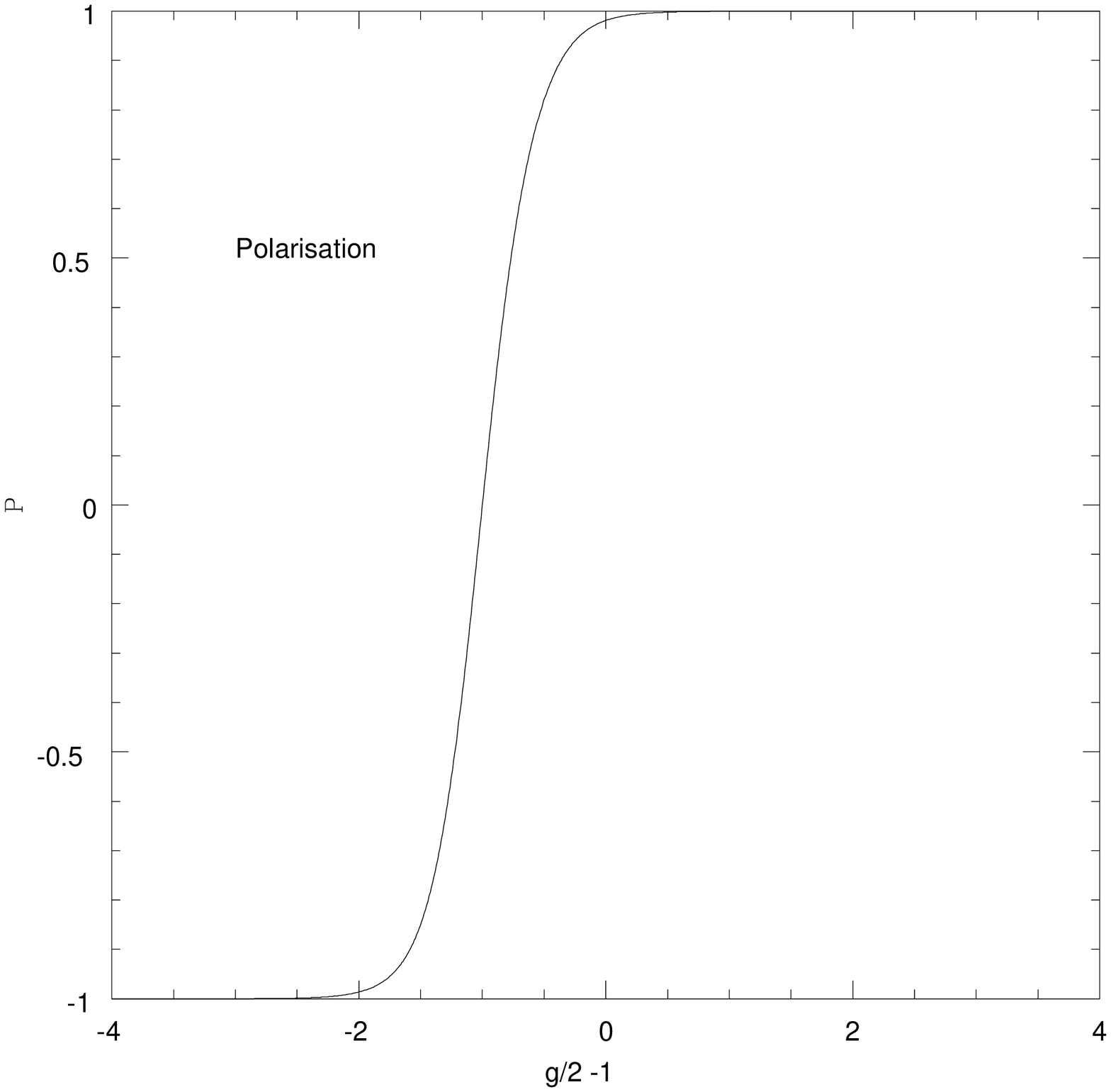}}
\centerline{Figure 5}
\centerline{\it Polarisation of the spin vs the g factor }
\bigskip

In which frame should one ask about the thermality of the spectrum of
fluctuations?  For an electromagnetic dipole moment, the damping is
proportional to ${D_{FW}^3 m\over D\tau^3 }- a^2 {D_{FW} m\over
D\tau}$ where $D_{FW}$ refers to the Fermi-Walker derivative of the
magnetic dipole moment $m$\cite{dipole}. Ie, the damping is related to
the rate of change of the magnetic moment in the Fermi-Walker frame.
This thus makes the Fermi Walker Frame the natural frame to look at the
fluctuations in. As argued above,
 the relevant correlation function for the spin polarisation is
stationary in all frames, including the Fermi-Walker frame. This thus
makes that frame the natural one to use to explain the results of the
spin polarisation.  Note that the $P_\uparrow -P_\downarrow$ is just
proportional to $(W+1)((W+1)^2+1) \Omega_0^3$, where $\Omega_0$ is the
precession frequency of the spin $\gamma^2\Lambda$, with $g$ factor 2
in a $B$ field of magnitude $B_0$. Since $g\Omega_0 $ is the precession
frequency in that $B$ field with g-factor of $g$, and since $\Omega_0$
 is also the acceleration
of the electron, this is just what we would expect from the radiation
reaction term if we replace the Fermi-Walker derivative with the
precession frequency $g\Omega_0$. But this is precisely what we would
expect by the fluctuation-dissipation theorem, as the fluctuations from
the pole at $t=0$ in the Wightman function is just the terms required
to maintain the commutation relations of the system against dissipation.

Thus in the Fermi-Walker frame, we can calculate the temperature of the
radiation bath by
\beq
T= -{g\Omega_0 \over\ln\left({P_\uparrow\over P_\downarrow}\right) }
\eeq
This temperature is frequency dependent (or since the frequency here is
proportional to $g$, is $g$ dependent), and is regular in this frame.
Figure 6 is a plot of the fluctuation Temperature seen by the spin vs
the g factor, with the Temperature plotted in units of the acceleration
$\gamma^2\Lambda$.   Note that the temperature in this case is slightly
more complex than it was in the scalar case, but has a very similar
form, namely it varies slightly (by about 20\%) across the frequency
range. The spin polarisation is exactly what one would expect naively
from the electron's coming into equilibrium with this temperature.

\epsfysize=3in
\centerline{\epsfbox{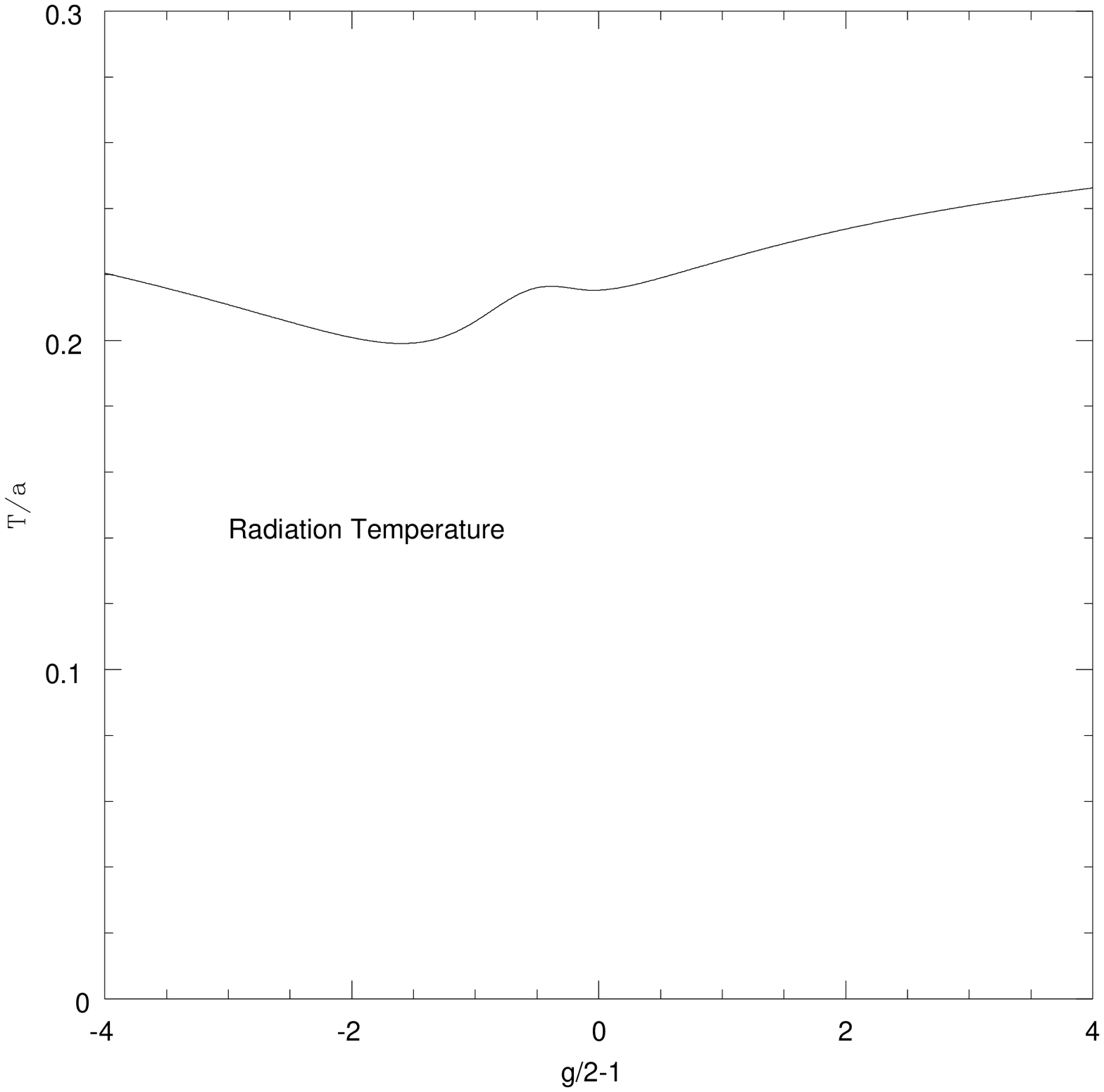}}
\centerline{Figure 6}
\centerline{\it The effective Temperature of the EM field in the Fermi-Walker Frame}
\bigskip

Note that if we had chosen the wrong frame, like the co-rotating frame, to try to calculate the effective temperature, we would have obtained a very different result for the temperature. The new effective temperature $T_{corot}= T_{FW} W/(W+1)$. Since $T_{FW}$ is regular with a temperature of about $.2 a$, the co-rotating temperature will diverge at $W=-1$ and go to zero at $W=0$. This effective temperature of the magnetic fluctuations in the co-rotating frame would thus be a wildly varying function of frequency,  going negative for $W$ between zero and one. This would be a key indication that one had chosen the wrong frame in which to calculate the effective temperature, even if the fluctuations are gaussian and stationary in that frame.

 Were the electron fixed in its orbit (ie, were the particle not
 allowed to deviate from the given orbit), then the above analysis would
be complete. However, for a real electron, the particle is not fixed in
its orbit, but can deviate therefrom. These excursions themselves can
be regarded as electric dipole moments around the fiducial circular
path, and the will themselves couple to the electromagnetic fields.
They will also couple to the spin. Let me begin by analysing these
dipole fluctuations in the absence of the spin and show that they
themselves also act as detectors for other components of the
electromagnetic field.

Let me assume that there is a Harmonic restoring force which drives the
particle back toward its fiducial orbit. This force will be assumed to
have a frequency $K$ and a damping term $\kappa$. The equation of
motion for these excursions in the various directions is therefore
\beq
\ddot X^i  +K^2 X^i= {e\over m} {\bf E}^i
\eeq 
where ${\bf E}^i$ has two components, the quantum fluctuations, and the
radiation reaction term. As I showed elsewhere\cite{dipole} ( and as BL
derived in a different manner), the radiation reaction term is 
\beq
E_{RR}^i ={2\over 3}({D^3_{FW}\over D\tau^3}-a^2 {D_{FW}\over D\tau}) eX^i
\eeq
since $eX^i$ is the dipole moment corresponding to this deviation from
the fiducial path. I will be interested in the fluctuations in the $z$
direction since those are the ones which will couple to the spin.  For
the $z$ direction, the FW derivative is the same as the ordinary one.
Assuming the radiation reaction force to be small, and that the time
derivatives in the radiation reaction term can be approximated by the
harmonic frequency, we have
\beq
\ddot Z +{2\over 3}(-{d^3\over d\tau^3}
+a^2 {d\over d\tau}) Z +K^2 Z= {e\over m}E_z
\eeq 
which has as solutions in the frequency domain for frequencies well
below the classical frequency (the speed of light divided by the
classical radius of the electron)
\beq
Z(\omega)=  {e\over m}{E _z(\omega) \over
 (-\omega^2 +i{2e^2\over3 m}\omega (\omega^2 +a^2) +K^2)}
\eeq
where $Z(\omega)= \int e^{-i\omega \tau} Z(\tau)d\tau$.
 Note that this equation has a pole near $-i3m/2e^2$ which corresponds to
the classical runaway solutions.  Since the damping term is only of
importance near the resonance ($\omega=K$),
 the $\omega^2+a^2$ term can be replaced
by $K^2+a^2$, to give a traditional damped oscillator equation with
damping coefficient
\beq
\kappa={2e^2\over3 m}(K^2+a^2)
\eeq
  Note that I have neglected the terms corresponding to the initial
  conditions for this z displacement, assuming that we have waited long
  enough for these to damp out.  For the oscillator, we can again
define raising and lowering operators in the traditional way
\beq
A = \sqrt{Km}Z +i P_z{1\over \sqrt{Km}}
\eeq
The transition probability for transitions to and 
from the $n^{th}$ states are given by
\ba
P_{\uparrow n}&&={1\over 2T}
( {e})^2\Sigma_{|1>}|\int<n+1|Z |n><1|E_z(\tau)|0>d\tau|^2 
\nonumber\\
&& = {e^2\over 2mK}(n+1)\int<E_x(\tau-i0)E_z(0)> e^{-iK\tau}d\tau
\\
&&= (n+1) {e^2\over 2mK}\left( \Theta(-K)({4\over 3} K(K^2+a^2)\right.\\ 
&& \left. \qquad \qquad \qquad +{1\over 72} e^{-2|K|\sqrt(3)/a } \left( 12\sqrt{3} K^2+30|K|a 
+13\sqrt{3} a^2\right)\right)
\nonumber\\
P_{\downarrow n}&&= P_{\uparrow n-1} + n{e^2\over 2mK}
({4\over 3} K(K^2+a^2))
\ea
Requiring detailed balance, so that the transitions
 between the states $|n>$ and $|n+1>$ balance requires
 that there be a population balance of
\beq
{N_{n+1}\over N_n}= {P_{\uparrow n}\over P_{\downarrow n+1}}
\eeq
which is independent of $n$. This gives an effective temperature of 
\beq
T= -{K\over \ln( {N_{n+1}\over N_n})}
\eeq
which I have plotted in figure 7 as a function of K. Again the system
sees a thermal spectrum with frequency dependent temperature which is
similar, but not identical to both that seen by the spin and that seen
by a scalar detector. (It is in fact more constant than that seen by a scalar detector).

\epsfysize=3in
\centerline{\epsfbox{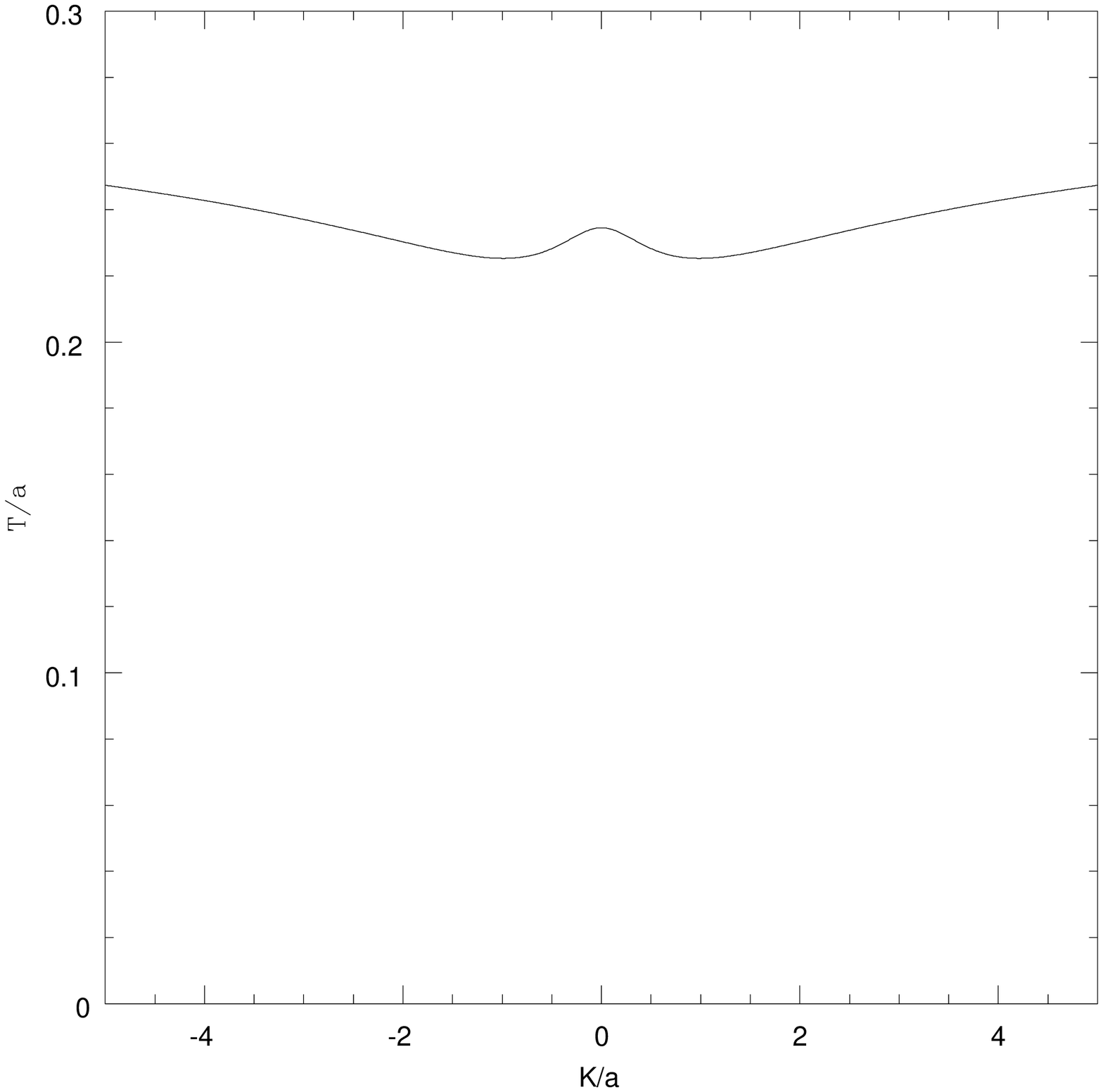}}
\centerline{Figure 7}
\centerline{\it The effective Temperature of the EM
 field affecting the oscillations in the z direction}

Finally we must take into account the coupling between these two
``detectors", namely the spin and the $z$ oscillations. Each separately
acts like a detector and sees a thermal spectrum of fluctuations
 of very similar
temperatures.   The $z$ oscillations  couple to the spin through two
routes. The first is if the classical magnetic field strength varies
in direction or amplitude  with position in the rest frame of the particle. The particle will then see a different field depending on where it is located. 
The second is an interaction through the velocity of the particle. This will
happen both because the particle will see the strong electric field
(causing the acceleration) as having a magnetic
 component due to any velocity of the electron from the fiducial
 circular orbit. Furthermore, the particle will have a different FW frame 
because of any such velocity, causing an extra "Thomas precession".   Variations in the magnetic field off the fiducial orbit are typically present,
 and I will assume that the magnetic field is arranged in the "weak
 focusing" configuration such that an effective restoring force to
 vertical oscillations is provided by the presence of off-orbit radial
 magnetic fields.

Let us look at the Thomas precession terms first.  The Fermi-Walker
transport equations are
\ba
{D_{FW}s^\mu\over Ds}&&= {Ds^\mu\over D\tau}
- V^\mu V_\alpha {Ds^\alpha\over D\tau}
\nonumber\\
&&=  {Ds^\mu\over Ds} + V^\mu a_\alpha s^\alpha
\ea
where $s$ is the path length along the path of the particle. I will work only to lowest order in the deviation, $Z$, from the path. We can define an orthonormal tetrad, based on the vectors $e^\mu_\alpha $ defined above by
\ba
&&\tilde e^\mu_t= V^\mu= e^\mu+ v_z e^\mu_z\\
&&\tilde e^\mu_\rho= e^\mu_\rho\\
&&\tilde e^\mu_\phi= e^\mu_\phi - \lambda v_z e^\mu_z\\
&&\tilde e^\mu_z= e^\mu_z + v_z e^\mu_t +\lambda v_z e^\mu_\phi
\ea
where $v_z= dZ/d\tau$, and $\lambda $ is an arbitrary constant.
Defining the components of the spin by
\beq
s_i=\tilde e^\mu_i s_\mu
\eeq
where $i$ is one of $\rho,\phi,z$, the derivatives of $s_i$ are given by
\beq
{ds_i\over ds}= {D_{FW}s_\mu\over Ds} e^\mu_i + s_\mu {D_{FW}e^\mu_i\over Ds}
\eeq
where $s$ is the path-length parameter along the path actually followed by the particle. But to lowest order in $v_z$, $ds=d\tau$. We then have
\ba
&&{ds_r\over d\tau}= {D_{FW}s_\mu\over d\tau}e^\mu_\rho+
 \gamma^2\Lambda\left[ s_\phi +(R\Lambda+\lambda) v_z s_z \right]\\
&&{ds_\phi\over d\tau}= {D_{FW}s_\mu\over d\tau}e^\mu_\phi
 -\gamma^2\Lambda s_r -\lambda \dot v_z s_z\\
&&{ds_z\over d\tau}= {D_{FW}s_\mu\over d\tau}e^\mu_z  
-\gamma^2\Lambda( R\Lambda +\lambda) s_\rho +\lambda \dot v_z s_\phi
\ea
But, $D_{FW}s_\mu/ds$ is just the local change of $s_\mu$ caused by the local ${\bf B}$ field. 

I will assume, as in weak focusing, that the magnetic field varies off the fiducial path, and in
 particular has 
 a non-zero radial component which varies with height ($z$). To
lowest order, in the rest frame of the ring, $B_\rho=-bz$. In the
fiducial rest frame of the electron, it sees both a magnetic field of
$\gamma bz$ in the $\rho$ direction and an electric field in the z
direction of $ -\gamma R\Lambda bz\approx -\gamma bz$. This electric
field will act as the harmonic restoring force for the electron's
excursions in the $z$ direction, giving a relation between the
restoring force of the oscillator and this field of
\beq
K^2= {e\over m}\gamma b 
\eeq

Assuming that $Z$ and the quantum fields $E,B$ are of the same order,  we have the local fields seen by the spin  are
\ba
&&{\bf B}_r={\bf F}_\mu\nu \tilde e^\mu_\phi \tilde e^\nu_z= B_r\\
&&{\bf B}_\phi={\bf F}_\mu\nu \tilde e^\mu_z \tilde e^\nu_r= -v_z(R\Lambda+\lambda) \gamma B_0 +B_\phi\\
&&{\bf B}_z={\bf F}_\mu\nu \tilde e^\mu_r \tilde e^\nu_\phi=\gamma B_0
\ea
which drive the Fermi-Walker changes of the spin.
The equations of motion can now be derived from the Hamiltonian
\ba
H= &&\left[ (g\mu_0 \gamma (B_0+B_z) -\gamma^2\Lambda  )s_z + g\mu_0(B_\phi s_\phi +B_\rho s_\rho)\right] \\
&&+\left\{  -\left(  v_z(RL+\lambda) (g\mu_0\gamma B_0-\gamma^2\Lambda)\right) s_\phi +(-\gamma bZ+\lambda \dot v_z) s_\rho \right\} 
\nonumber
\ea
where the term in the square brackets is the term we have already 
used for the spin travelling along the fiducial path, and the 
terms in the curly brackets,
$H_I$, are  caused by the motion away from the fiducial path. 

I will use this Hamiltonian to calculate the effects of the deviations from the  path from the fiducial path. In doing so I will take $\lambda=0$. As I will argue below, the results are independent of $\lambda$, at least to the order to which I am carrying out the calculation.

  For the perturbations
on the spin system, write the above equations in terms of
 $\sigma_+$ and $\sigma_-$. Using $\gamma\Lambda = 2\mu_0 B_0 $ and $\Omega= \mu_0(g-2)\gamma B_0$, and dropping the $B_zs_z$ term as not contributing to the spin flip transition rate,  I define
\beq
{\cal Z}_{\pm} = \half (\pm i\dot Z\Omega + {g\over 2} K^2 Z)
\eeq
to give
\beq
H_I=-{\cal Z}_- \sigma_+ -{\cal Z}_+ \sigma_-
\eeq

If we were to assume that the coupling to $B$ field 
fluctuations by the spin 
are zero, than these cross couplings from the motion
of the harmonic oscillators would give transitions of the form
\ba
P_{I\uparrow}= < {\cal Z}_+(\tau-i0){\cal Z}_-> e^{-i\Omega\tau} d\tau\\
P_{I\downarrow}= < {\cal Z}_+(\tau+i0)  {\cal Z}_-> e^{-i\Omega\tau} d\tau
\ea
 I get
\ba
P_{I\uparrow}&&=\mu_0^2{ (\Omega^2 
- {g\over 2} K^2)^2 \over  (\Omega^2 -K^2)^2+\Omega^2\kappa^2  } \int <E_z(\tau-i0)E_z(0)>e^{-i\Omega\tau}d\tau\\
 &&=\mu_0^2(\gamma^6\Lambda^3)  {(\Omega^2 - {g\over 2} K^2)^2 \over  (\Omega^2 -K^2)^2+\Omega^2\kappa^2  }(-G_0\Theta(-W) + e^{-2\sqrt{3} |W|} G_1)
\nonumber\\
P_{I\downarrow}&&= P_{I\uparrow} +\mu_0^2(\gamma^6\Lambda^3)  
{(\Omega^2 - {g\over 2} K^2)^2 \over  (\Omega^2 -K^2)^2+\Omega^2\kappa^2  }
\ea
where 
\ba
G_0&&= {4\over 3} W(W^2+1)\\
G_1&&= {1\over 72 } \left( \sqrt{3}(12 W^2+13)+30|W|)\right)
\nonumber
\ea

However there are correlations between the
 components $E_z$ which drive the vertical oscillations,  and the $B_\pm$ components which drive the spin flips.
  These add additional 
complications to the expressions. Let me define these correlation spin flip probabilities by
\ba
P_{C\uparrow}&&= -g\mu_0\left( \int( <B_+(\tau-i0){\cal Z}_-(0)>
+<{\cal Z}_+(\tau-i0)B_-(0)>)e^{-i\Omega\tau} d\tau\right)\\
&&=-g\mu_0^2\left( ({g\over 2} K^2-\Omega^2)
\int\left( {<B_+(\tau-i0)E_z(0)> \over (K^2+i\Omega\kappa-\Omega^2)}
   +{<E_z(\tau-i0) B_-(0)>\over(K^2-i\Omega\kappa-\Omega^2) }\right)
 e^{-i\Omega\tau}d\tau\right) 
\nonumber \\
&&\approx -g\mu_0^2 \left({{g\over 2}K^2
-\Omega^2\over K^2-\Omega^2}\right)
(\gamma^2\Lambda)^3\left(-\Theta(-W) L_0 
+e^{-2\sqrt{3}|W|}(L_1 +{|W|\over W} L2)\right) \nonumber
\ea
where I have have assumed $\kappa$ is very small.
  Here we have
\ba
&&L_0=-{1\over 3}(2W+1) \nonumber\\
&&L_1= -{\sqrt{3}\over 24}(4W^2-5W+3)\\
&&L_2= {1\over 24}(6W^2 -10W +4)  \nonumber
\ea

The total interactive transition rate is then
\beq
P_{T\uparrow}= P_{\uparrow} +P_{C\uparrow}+P_{I\uparrow}
\eeq
and similarly for 
$P_{T\downarrow}$. These lead to the polarisation curve of figure 1.

There are a number of points which must be emphasised.
 The first is that it is the transverse oscillations of the electron
which drive the extra terms
 in the transition. In the above I assumed that the only effects which
drive those transverse oscillations are the vacuum fluctuations of the
electric field in the rest frame of the particle. If there are other
sources of noise which drive those transverse oscillations (field
misalignments, interactions with other particles in the beam, etc) then
those will all also contribute to the spin polarisation.  (Note that
the action of the spin system on the oscillator is not a source of
significant noise. The coupling between the spin and the Harmonic
oscillator--reduced to dimensionless variables-- is of order
$\sqrt{\hbar K\over mc^2}(K\pm \Omega)$ which is clearly tiny. The only
reason the oscillator effects the spin is that its coupling to the
electric field in the same units is enhanced by the factor
$\sqrt{mc^2\over \hbar K}$ over the direct coupling between the magnetic fluctuations and the spin, compensating for the weakness of the
coupling between the two systems).

Secondly, for $g=2$ the factor 
$gK^2/2-\Omega^2\over K^2\pm i\kappa\Omega -\Omega^2$ is unity
 except at the resonance. However for $g$ not equal to 2, the 
system behaves very differently below the resonance than above.

In the co-rotating frame, the   transition rates caused by the direct coupling to the local magnetic field, and the transition rate induced by the motion
of the electron have very different
forms. The direct transition rate is centered at g=0 (figure 5)
while those for the oscillator are centered at g=2. The final
polarisation curve is thus a mixture of the two, resulting in the shift
to about $g=.8$ of figure 1. Also, the relative weights of the two
terms shift with frequency, giving the fine structure in the figure 1.

I have argued that both detectors are immersed in essentially identical
temperature heat baths. Why would the interaction not then not bring
the joint system to the same joint temperature? The important point
here is that    the Fermi-Walker frame is the natural frame for the
spin, and is the frame in which the fluctuations look thermal.
However, the interaction is static in the co-rotating frame.  The $z$
oscillations and $E_z$ are unaffected by the transformation to the FW
frame, and thus in this frame both systems independently have the same
distribution. However, in that frame, the interaction between the
oscillator and the heat bath is no longer static, but   is time
dependent.  (although $Z$ is unaffected by the transformation, the
coupling is through $\sigma_\rho$ and $\sigma_\theta$ which {\bf do}
change under the transformation.)
  
 Because of this time dependent coupling,   the harmonic system tends
to preferentially drive
 up-transitions in the spin system (and the spin system preferentially
drives down-transitions in the oscillator). The  coupling acts like an
amplifier, driving the spin system to its higher energy state (at least
for   $g>0$). The power for this amplifier clearly comes from the
motion around the storage ring.

 The electron system thus acts   like the snow on the TV screen.
 Without understanding the details of the system, one cannot conclude
that the lack of thermal output (the spin's polarisation is not
thermally distributed) in the detector demonstrates a lack of thermal
input (the fluctuations in the relevant components of the EM field in
the FW frame).

In both DK and Jackson, the claim appears to be that the standard
result is obtained for calculating the spin flip by the fluctuations
with the electron assumed to be travelling along the fiducial circular
path (see eqn 44-47 in Jackson and the first equation in section 5 for
DK).  However in both cases they assume that the fluctuating quantum
fields cause a Thomas precession. They use the Thomas-BMT equations for
the evolution of the spin, which assume that the fields drive the
accelerations away from the fiducial path which drives the Thomas
precession. However, this extra Thomas precession,
caused by the quantum fields, is a kinematic effect, and arises soley
because of   deviations in the electron's orbit.   Neither Jackson nor
DK explicitely examines the behaviour of the
electron deviations from the fiducial orbit in computing the standard
(figure 1) depolarisation for an electron. Their
 calculations thus effectively assume that $K=0$and that $\kappa=0$.

Also, in my derivation  above, I set $\lambda=0$ which meant that the
Thomas precession terms due to the $z$ motion depended only on the
velocity in the $z$ direction.  On the other hand, by using the TBMT
equations, Jackson and DK have
 an acceleration dependent term in their equations (which they then set
 equal to the forces exerted on the particle by the vacuum fields to
get the TBMT equations).  They were thus implicitely taking $\lambda
=-R\Lambda$, which makes the Thomas precession  terms depend only on the $z$
acceleration.  However, if one kept the $\lambda$ dependent terms in
$H_I$, this would add
terms of the form $\lambda ( (\pm i\Omega \dot Z - +\ddot Z)$ to ${\cal
Z}_\pm$.  But ${\cal Z}_{\pm}$ enters the equations for the spin flip
probabilities integrated with $e^{-i(\pm\Omega\tau)}$, which transforms
$\ddot Z$ to $i\Omega \dot Z$ and converts the $\lambda $ dependent
terms in ${\cal Z}_{\pm}$ to zero.

\section{Emission of radiation by accelerated particle}

Unruh and Wald \cite{UW} examined the behaviour of a detector under
uniform acceleration, and in particular the emission of radiation by
such a detector. We looked at two cases in particular. In the one
case we asked for the state of the field under the condition that a
two level detector was found at the end of the process to have been
excited. In this case they found that the field was in a single
particle excited state. That single particle was concentrated in the
region which did not have causal contact with the detector. In the case
where the detector was found after the experiment to be unexcited, the
field was in a coherent superposition of an unexcited vacuum state, and
a two particle excited state.

Here I will examine the question from a slightly different point of
view. The detector is taken to be a harmonic oscillator as above. I
will ask what the state is of the radiation field (the massless scalar
field) without any measurement of the state of the detector.

The equation of motion for an internal oscillator coupled to a scalar field is
\ba
\ddot q +2\epsilon^2 q +\Omega^2 q= \epsilon\phi_0
\\
\partial_t^2\phi -\nabla^2\phi 
=\epsilon \int q(\tau)\delta^4(x^\mu-X^\mu(\tau))d\tau
\ea
where $X^\mu(\tau)$ are the coordinates of
 the particle at proper time $\tau$, and 
\beq
\phi=\phi_0 +\int {  G}(x^\mu -X^\mu(\tau))d\tau
\eeq
with ${  G}$ the retarded Green's function. The solution for $q$ is
\beq
q(t)= \epsilon\int^\tau 
{\sin(\tilde \Omega(\tau-\tau'))\over \tilde \Omega }
e^{\lambda(\tau-\tau')} \phi_0(\tau') d\tau'
\eeq
while the solution for the field $\phi$ is
\beq
\phi(t,x)= \phi_0(t,x)
 + \epsilon \int q(\tau)\Theta(t-T(\tau))\delta( (t-T(\tau))^2-(x-X(\tau))^2) d\tau
\eeq
Ie, the final outgoing field is a linear function of the ingoing field.
This implies that the annihilation and creation operators or the
outgoing field are linear combinations of those for the ingoing field.
If the outgoing annihilation operators were functions only of the
ingoing annihilation operators, which is what happens when the particle
is unaccelerated, then the state of the outgoing field will be
indistinguishable from the vacuum state. On the other hand, if the
outgoing field's ($\phi$'s) annihilation operators are function of both
the ingoing field's ($\phi_0$'s) annihilation and creation operators,
then that outgoing field will in general by a ``squeezed state" with
respect to the outgoing vacuum state.  A squeezed state has the property
that it is a coherent sum of even numbers of particle states. Ie, the
detector ``scatters" the vacuum fluctuations in the field $\phi_0$ so
as to produce correlated pairs of particles in the outgoing state.
Since this state is again a gaussian state, it is completely
characterised by the pairwise correlation function
$C=<\phi(t,x)\phi(t',x')>$ . We thus have
 
\ba
C((t,x)&&,(t',x')) = <\phi_0(t,x)\phi_0(t',x')> 
+\epsilon^2 \int G((t',x'),\tau') <\phi_0(t,x)\phi_0(\tau')> 
\nonumber\\
&& +\epsilon^2 \int G((t,x),\tau') <\phi_0(\tau')\phi_0(t',x')> 
\nonumber\\
&&+ \epsilon^4 \int\int G((t,x),\tau')G((t',x'),\tau")
<\phi_0(\tau')\phi_0(\tau")>d\tau d\tau'
\ea
where 
\beq
G((t,x),\tau')=  \int_\tau'  
{\sin(\tilde \Omega(\tau-\tau'))\over \tilde \Omega }
 e^{-\lambda(\tau-\tau')} \phi_0(\tau') 
 \Theta(t-T(\tau))\delta( (t-T(\tau))^2-(x-X(\tau))^2) d\tau
\eeq
Ie, the two point function with the detector can be calculated as a
function of the two point function of the free field.  However,
measuring these correlations would be difficult, since in general the
two correlated particles are separated by a time of order the decay
time of the system. For any known accelerating system (eg electrons in
a synchrotron) that time scale is of the order minutes or hours.

 \section*{Acknowledgements}
I would like to thank P. Chen and David Cline for inviting me to their
respective conferences and reviving my interest in this subject. I
would like to thank David Jackson whose doubts about treating the
circular acceleration as a thermal bath incited me to look more closely
at that problem. Finally I thank the CIAR and NSERC for support while
this work was being done.

\end{document}